\documentclass[12pt,a4paper]{article}
\usepackage{myart}
\usepackage{amsmath}
\usepackage{graphicx}
\usepackage{amsfonts}
\usepackage{amssymb}

\begin{document}
\hfill\hbox{LPM-01-30}

\hfill\hbox{September, 2001}

\bigskip

\begin{center}
{\Large \textbf{Scaling Lee-Yang Model on a Sphere.}}

{\Large \textbf{I. Partition Function}}\footnote{Extended version of a talk
presented at the NATO Advanced Research Workshop on Statistical Field
Theories, Como 18--23 June 2001.}

\vspace{1.5cm}

{\large Al.Zamolodchikov}\footnote{On leave of absence from Institute of
Theoretical and Experimental Physics, B.Cheremushkinskaya 25, 117259 Moscow, Russia}

\vspace{0.2cm}

Laboratoire de Physique Math\'ematique\footnote{Laboratoire Associ\'e au CNRS URA-768}

Universit\'e Montpellier II

Pl.E.Bataillon, 34095 Montpellier, France
\end{center}

\vspace{1.0cm}

\textbf{Abstract}

Some general properties of perturbed (rational) CFT in the background metric
of symmetric 2D sphere of radius $R$ are discussed, including conformal
perturbation theory for the partition function and the large $R$ asymptotic.
The truncated conformal space scheme is adopted to treat numerically perturbed
rational CFT's in the spherical background. Numerical results obtained for the
scaling Lee-Yang model lead to the conclusion that the partition function is
an entire function of the coupling constant. Exploiting this analytic
structure we are able to describe rather precisely the ``experimental''
truncated space data, including even the large $R$ behavior, starting only
with the CFT information and few first terms of conformal perturbation theory.

\section{Motivations}

There are several reasons to be interested in euclidean 2D field theory in
spherical background. Apart from conceptual interest in relativistic field
theory on curved spaces, the particular example of sphere presents an apparent
simplification of a symmetric space where physics is independent on the space
position while being different from that on the infinite flat plane.

It has been proven useful to study properties of interacting field theory
through finite-size effects (see e.g.\cite{Luscher, TBA, YuZ}). The compact
geometry of sphere naturally provides a settlement of finite size problem,
different from the Casimir effect which has been mainly studied so far.

Exact constructions of various conformal field theories (CFT) in 2D gave rise
to a very fruitful idea to consider non-conformal (both massive and massless)
field theory models as certain perturbations of CFT's by relevant operators
\cite{Zam1}. One considers formal actions like
\begin{equation}
\mathcal{A}_{\lambda}=\mathcal{A}_{\text{CFT}}+\frac\lambda{2\pi}\int
\phi(x)d^{2}x\label{Alambda}%
\end{equation}
where $\phi$ is some relevant primary field in the theory of dimension
$\Delta$ and $\lambda$ is the corresponding coupling constant. Even if such
perturbation leads to a field theory which is not integrable, the perturbation
theory in $\lambda$, called the conformal perturbation theory (CPT), may give
valuable information about the short distance behavior of the model. One of
the main problems of CPT is related to infrared divergences of the
perturbative integrals, which result in certain non-analyticity in $\lambda$.
Finite geometry of a sphere imposes an infrared cutoff. It is natural to
expect that on a finite sphere observables are analytic in $\lambda$ at
$\lambda=0$. Below I'll try to argue that in certain cases the CPT series on a
finite sphere is \emph{absolutely convergent} and therefore results in entire
function of $\lambda$. Although there are no obvious reasons to expect any
sort of integrability on the sphere even if the model (\ref{Alambda}) is
integrable on the flat plane, the analytic property suggested is extremely
restrictive. As we'll see before long, it permits to get a lot of information
starting from few first CPT coefficients and some general information about
$\lambda\rightarrow\infty$ asymptotic.

There is an approximate way to treat the actions like (\ref{Alambda}) called
the truncated conformal space (TCS) approach \cite{YuZ}. Sometimes it gives
rather precise numerical information even about the large-distance properties
of the field theory \cite{TCS}. Spherical geometry seems to be a very natural
environment for the TCS approach.

The main motivation, at least for me, to study perturbed CFT's on fixed sphere
is the long standing challenge of 2D quantum gravity. There are all reasons to
believe that the field theoretic approach based on the Liouville field theory
is a relevant way to understand 2D gravity at least in the so-called ``week
coupling regime''. However, up to now only very basic information about
general scale dimensions and simplest correlation functions is reached by the
field theoretic means, even in the simplest case of spherical topology. This
has to be compared with very reach ``experimental'' results coming from the
matrix models and similar ``discrete geometry'' approaches (see e.g.\cite{KKK}
for most recent achievements in the field). I hope that better understanding
of the field theory on a fixed symmetric sphere can be a starting point to
approach the problem at least in the semiclassical limit where the central
charge of the massless matter modes is large negative and the geometry becomes ``rigid''.

\section{CFT in curved background}

The most advanced understanding of the 2D field theory in curved background
$g_{ab}(x)$ is reached in conformal field theory. This is due to the following
two simple properties postulated as a basis of CFT construction

\textbf{Stress tensor anomaly.} In 2D CFT the trace of the stress tensor
$T_{ab}(x)$ is in fact a c-number (i.e., proportional to the identity
operator) and reads explicitly
\begin{equation}
\theta(x)=g^{ab}T_{ab}=-\frac c{12}\widehat{R}\label{trace}%
\end{equation}
where $\widehat{R}$ is the scalar curvature of the background metric and $c$
is certain (real) number, called the central charge. The central charge is the
basic characteristic for any particular CFT.

\textbf{Primary fields. }Among the set of local fields $\{\Phi\}$ in the
theory there is a subset of so called primary fields $\phi_{i}$ which
transform very simply (here the primary fields are supposed to be scalars)
\begin{equation}
\delta\phi_{i}(x)=-\Delta_{i}\phi_{i}(x)\delta\varphi(x)\label{primary}%
\end{equation}
under the Weil variations of the metric
\begin{equation}
\delta g_{ab}(x)=g_{ab}(x)\delta\varphi(x)\label{Weil}%
\end{equation}
In (\ref{primary}) $\Delta_{i}$ are (real) numbers characteristic for the
primary fields, called the scale dimensions (or conformal weights). All other
local fields can be found in the operator product expansions of primaries and
the nontrivial stress tensor components. Especially simple are the so called
rational CFT's which involve only finite number of primary fields (and
therefore contain finite spectrum of dimensions $\Delta_{i}$).

Conceptually these two properties (which were in fact abstracted from explicit
calculations in certain simple examples like free field theories) are enough
to develop the whole structure as rich as the conformal field theory.

In 2D one can always choose locally an ``isothermal'' coordinate system where
the metric is ``conformally flat''
\begin{equation}
g_{ab}(x)=e^{\varphi(x)}\delta_{ab}\label{isothermal}%
\end{equation}
In the case of sphere such coordinates $(z,\bar z)$ can be chosen almost
globally, covering all the surface except for the ``south pole'' $z=\infty$.
Simple transformation low (\ref{primary}) allows to express any correlation
function of primary fields through the correlation functions on infinite flat
plane with $g_{ab}=\delta_{ab}$ (again all fields are implied scalar)
\begin{equation}
\left\langle \phi_{1}(x_{1})\ldots\phi_{n}(x_{n})\right\rangle _{\varphi
}=\prod_{i=1}^{n}e^{-\Delta_{i}\varphi(x_{i})}\left\langle \phi_{1}%
(x_{1})\ldots\phi_{n}(x_{n})\right\rangle _{\varphi=0}\label{flatten}%
\end{equation}

In this paper we'll concentrate on the metric of sphere of radius $R$. The
Weil factor $\exp\varphi(z,\bar z)$ in (\ref{isothermal}) can be chosen in the
form
\begin{equation}
e^{\varphi(z,\bar z)}=\frac{4R^{2}}{(1+z\bar z)^{2}}\label{sphere}%
\end{equation}
The scalar curvature is a positive constant
\begin{equation}
\widehat{R}=-4e^{-\varphi}\partial\bar\partial\varphi=2R^{-2}\label{curvature}%
\end{equation}
Metric (\ref{sphere}) is invariant under the $O(3)$ group of coordinate
transformations
\begin{equation}
z\rightarrow\frac{az+b}{\bar a-\bar bz}\label{O(3)}%
\end{equation}
where it is convenient to set $a\bar a+b\bar b=1$.

The stress tensor anomaly (\ref{trace}) readily prescribes the $R$ dependence
of the partition function
\begin{equation}
Z_{\text{CFT}}(R)=R^{c/3}Z_{0}\label{ZR}%
\end{equation}
Here $Z_{0}=Z_{\text{CFT}}(1)$ is the CFT partition function at $R=1$. As long
as I know for this multiplier any theoretical predictions are lacking. In what
follows I'll consider only the ratio $Z_{\text{CFT}}(R)/Z_{0}$. A simple
consequence of eq.(\ref{Weil}) is the $R$ dependence of any correlation
function of primary fields on the sphere
\begin{equation}
\left\langle \phi_{1}(x_{1})\ldots\phi_{n}(x_{n})\right\rangle _{R}%
=R^{-2\sum_{i=1}^{n}\Delta_{i}}\left\langle \phi_{1}(x_{1})\ldots\phi
_{n}(x_{n})\right\rangle _{R=1}\label{phiR}%
\end{equation}

\section{Perturbation}

In non-trivial background (\ref{isothermal}) perturbed action (\ref{Alambda})
is replaced by
\begin{equation}
\mathcal{A}_{\lambda}=\mathcal{A}_{\text{CFT}}+\frac\lambda{2\pi}\int
\phi(x)e^{\varphi(x)}d^{2}x\label{cpert}%
\end{equation}
where $\phi$ is again a relevant primary field of dimension $\Delta<1$.

Formal expression (\ref{cpert}) gives rise to perturbative expansion in
$\lambda$. Consider e.g., the perturbed partition function $Z_{\lambda
}[\varphi]$. Formally
\begin{equation}
\frac{Z_{\lambda}[\varphi]}{Z_{\text{CFT}}[\varphi]}=\sum_{k=0}^{\infty}%
\frac{(-\lambda)^{k}}{(2\pi)^{k}k!}\int\left\langle \phi(x_{1})\ldots
\phi(x_{n})\right\rangle _{\varphi}\prod_{i=1}^{n}e^{\varphi(x_{i})}d^{2}%
x_{i}\label{Zlambda1}%
\end{equation}
where the correlation functions are computed in unperturbed CFT but in
non-trivial background $\varphi$. With the use of (\ref{flatten}) one can
``flatten'' these correlation functions, reducing (\ref{Zlambda1}) to an
expression which involves only the CFT correlation in trivial flat background
$\varphi=0$%
\begin{equation}
\frac{Z_{\lambda}[\varphi]}{Z_{\text{CFT}}[\varphi]}=\sum_{k=0}^{\infty}%
\frac{(-\lambda)^{k}}{(2\pi)^{k}k!}\int\left\langle \phi(x_{1})\ldots
\phi(x_{n})\right\rangle _{\text{CFT}}\prod_{i=1}^{n}e^{(1-\Delta
)\varphi(x_{i})}d^{2}x_{i}\label{Zlambda2}%
\end{equation}
On the sphere (\ref{sphere}) the $R$ dependence is summarized as follows
\begin{equation}
\frac{Z_{\lambda}(R)}{Z_{0}R^{c/3}}=z(h)\label{zh}%
\end{equation}
where the dimensionless combination
\begin{equation}
h=\lambda(2R)^{2-2\Delta}\label{h}%
\end{equation}
is introduced. Eq.(\ref{Zlambda2}) gives $z(h)$ as a regular power series in
$h$
\begin{equation}
z(h)=\sum_{n=0}^{\infty}(-h)^{n}z_{n}\label{zn}%
\end{equation}
where explicitly
\begin{align}
z_{0} &  =1\nonumber\\
z_{n} &  =\frac1{(2\pi)^{n}n!}\int\left\langle \phi(x_{1})\ldots\phi
(x_{n})\right\rangle _{\text{CFT}}\prod_{i=1}^{n}\frac{d^{2}x_{i}}%
{(1+x_{i}\bar x_{i})^{2-2\Delta}}\label{zint}\\
\  &  =\frac\pi{(2\pi)^{n}n!}\int\left\langle \phi(0)\phi(y_{2})\ldots
\phi(y_{n})\right\rangle _{\text{CFT}}\prod_{i=2}^{n}\frac{d^{2}y_{i}%
}{(1+y_{i}\bar y_{i})^{2-2\Delta}}\;\;\;\;\text{at }\;\;\text{ }n>0\nonumber
\end{align}
In the last equation the symmetry (\ref{O(3)}) is used to eliminate one of the
integrations. In CFT two- and three-point functions are fixed up to
normalization constants. Assuming the usual in CFT normalization of primary
fields we have
\begin{align}
\left\langle \phi(x)\right\rangle _{\text{CFT}} &  =0\nonumber\\
\left\langle \phi(x_{1})\phi(x_{2})\right\rangle _{\text{CFT}} &
=\frac1{\left|  x_{1}-x_{2}\right|  ^{4\Delta}}\label{corr123}\\
\left\langle \phi^{\text{(c)}}(x_{1})\phi^{\text{(c)}}(x_{2})\phi^{\text{(c)}%
}(x_{3})\right\rangle _{\text{CFT}} &  =\frac{C_{\phi\phi\phi}}{\left|
(x_{1}-x_{2})(x_{2}-x_{3})(x_{3}-x_{1})\right|  ^{2\Delta}}\nonumber
\end{align}
where $C_{\phi\phi\phi}$ is the corresponding structure constant, which is
explicitly known in exact CFT constructions. First two coefficients in
(\ref{zn}) are readily calculated
\begin{align}
z_{1} &  =0\label{z2}\\
z_{2} &  =\frac1{8(1-2\Delta)}\nonumber
\end{align}
while for $z_{3}$ we have
\begin{align}
z_{3} &  =\frac{\pi C_{\phi\phi\phi}}{6(2\pi)^{3}}\int\frac{d^{2}y_{1}%
d^{2}y_{2}}{[(1+y_{1}\bar y_{1})(1+y_{2}\bar y_{2})]^{2-2\Delta}[y_{1}\bar
y_{1}y_{2}\bar y_{2}(y_{1}-y_{2})(\bar y_{1}-\bar y_{2})]^{\Delta}}%
\label{z3}\\
\  &  =\frac{C_{\phi\phi\phi}}{24}\frac{\Gamma(2-3\Delta)\Gamma(2-\Delta
)}{\Gamma(4-4\Delta)}\int_{0}^{1}v^{-\Delta}dvF(\Delta,\Delta,1,v)F(2-2\Delta
,2-3\Delta,4-4\Delta,1-v)\nonumber
\end{align}
Higher perturbative integrals in (\ref{zint}) involve four and more point CFT
correlation functions. These are not that universal as (\ref{corr123}) and
require more detailed information about a particular CFT model.

If the series (\ref{zn}) is convergent, it defines the perturbed partition
function as a function of $\dot R$. In what follows we'll see an evidence that
in certain cases this series is indeed convergent and, moreover, absolutely
convergent, so that $z(h)$ is an entire function of $h.$

In any case the perturbative development (\ref{zn}) describes the
$R\rightarrow0$ asymptotic of the partition function $Z_{\lambda}(R)/Z_{0}$.
Let's now turn to the opposite $R\rightarrow\infty$, or $h\rightarrow\infty$ limit.

\section{Large $R$ asymptotic}

It seems intuitively clear that when $R\gg m^{-1}$ (where $m\sim
\lambda^{1/(2-2\Delta)}$ is the typical mass scale in the perturbed model) the
local physics is almost the same as in infinite flat space-time. In particular
the leading exponential asymptotic of the partition function $Z_{\lambda}(R)$
is expected to be governed by the specific vacuum energy $\mathcal{E}%
_{\text{vac}}$ in flat space
\begin{equation}
\log Z_{\lambda}(R)\sim-4\pi R^{2}\mathcal{E}_{\text{vac}}+\ldots
\label{leading}%
\end{equation}
This vacuum energy is an important non-perturbative characteristic of a field
theory model. For dimensional reasons
\begin{equation}
\mathcal{E}_{\text{vac}}=-A\lambda^{1/(1-\Delta)}\label{Evac}%
\end{equation}
where $A$ is a dimensionless number. In integrable field theory this number is
typically known exactly (see e.g., the example below).

To get an idea about further $R\rightarrow\infty$ corrections let's start with
the relation
\begin{equation}
\frac{d\log Z_{\lambda}(R)}{dR^{2}}=-\left\langle \theta\right\rangle
\label{theta}%
\end{equation}
and take the rescaled metric
\begin{equation}
e^{\varphi}=\left(  1+\frac{z\bar z}{4R^{2}}\right)  ^{-2}\label{rescaled}%
\end{equation}
As $R\rightarrow\infty$ this Weil factor is trivial $e^{\varphi}=1$ and
$\left\langle \theta\right\rangle $ coincides with the expectation value on
infinite flat plane
\begin{equation}
\left\langle \theta\right\rangle _{\text{flat}}=4\pi\mathcal{E}_{\text{vac}%
}\label{tflat}%
\end{equation}
reproducing the leading asymptotic (\ref{leading}). For next corrections we
should take into account the stress tensor anomaly (\ref{trace})
\begin{equation}
\theta=-\frac c{6R^{2}}+\theta_{\text{flat}}\label{tc}%
\end{equation}
as well as the deviation of the metric from flat around the location of the
operator $\theta(x)$. Let's put $x=0$ so that
\begin{equation}
e^{\varphi}=1-\frac{z\bar z}{2R^{2}}+\frac{3(z\bar z)^{2}}{16R^{4}}%
+\ldots\label{phix}%
\end{equation}
General variation formula ($X$ is any observable)
\begin{equation}
\delta\left\langle X\right\rangle =-\frac1{4\pi}\int\left\langle
\theta(x)X\right\rangle e^{\varphi(x)}\delta\varphi(x)d^{2}x\label{genvar}%
\end{equation}
to the first order in $1/R^{2}$ gives rise to the following $R^{-2}$
correction to $\left\langle \theta\right\rangle $
\begin{equation}
\left\langle \theta(0)\right\rangle _{\text{sphere}}=\left\langle
\theta(0)\right\rangle _{\text{flat}}+\frac1{8\pi R^{2}}\int\left\langle
\theta(x)\theta(0)\right\rangle _{\text{flat}}x^{2}d^{2}x-\frac c{6R^{2}%
}+O(R^{-4})\label{tvar}%
\end{equation}
The famous $c$-theorem sum rule \cite{csum} allows to evaluate the integral
\begin{equation}
\int\left\langle \theta(x)\theta(0)\right\rangle _{\text{flat}}x^{2}%
d^{2}x=\frac{4\pi c}3\label{ct}%
\end{equation}
The anomalous $R^{-2}$ term is cancelled out and the IR\ corrections to
$\left\langle \theta\right\rangle $ appear at the order $R^{-4}$, so that
\begin{equation}
\left\langle \theta(0)\right\rangle _{\text{sphere}}\sim4\pi\mathcal{E}%
_{\text{vac}}+\frac{b_{1}}{R^{4}}+\frac{2b_{2}}{R^{6}}+\ldots\label{tcorr}%
\end{equation}
with some (dimensional) coefficients $b_{1}$, $b_{2}$, $\ldots$.Thus, as it
can be intuitively expected, at $R\rightarrow\infty$ there is no power
correction to $Z_{\lambda}(R)$ and the IR expansion has the form
\begin{equation}
\log\frac{Z_{\lambda}(R)}{Z_{0}}\sim-4\pi R^{2}\mathcal{E}_{\text{vac}}%
+\log(z_{\infty})+\frac{b_{1}}{R^{2}}+\frac{b_{2}}{R^{4}}+\ldots\label{ZIR}%
\end{equation}
Note that the sum rule (\ref{ct}) plays essential role in the cancellation of
the $\log R$ correction. The (dimensional) integration constant
\begin{equation}
Z_{\infty}=\lambda^{-c/(6-6\Delta)}z_{\infty}\label{zinf}%
\end{equation}
where $z_{\infty}$ is a dimensionless number. Presently I don't understand
neither physical meaning of this constant nor how it can be predicted on the
basis of flat field theory. Nevertheless this parameter seems to be an
important characteristic of the perturbed CFT on a sphere.

In terms of the variable (\ref{h}) asymptotic (\ref{ZIR}) reads
\begin{equation}
\log z(h)=\pi Ah^{1/(1-\Delta)}+\log(2^{c/3}z_{\infty})-\frac c{6-6\Delta}\log
h+a_{1}h^{-1/(1-\Delta)}+a_{2}h^{-2/(1-\Delta)}+\ldots\label{zIR}%
\end{equation}
where $A$ is defined in (\ref{Evac}) and $a_{1}=4b_{1}\lambda^{1/(1-\Delta)}$,
$a_{2}=16b_{2}\lambda^{2/(1-\Delta)}$ etc.

Unlike $z_{\infty}$, coefficients $a_{1}$, $a_{2}$, etc., (or $b_{1}%
,b_{2},\ldots$ in (\ref{ZIR})) can be expressed in terms of higher correlation
functions of $\theta$ in flat space-time. For example
\begin{equation}
b_{1}=\frac1{128\pi^{2}}\int\left\langle \theta(x_{1})\theta(x_{2}%
)\theta(0)\right\rangle _{\text{flat}}x_{1}^{2}x_{2}^{2}d^{2}x_{1}d^{2}%
x_{2}-\frac3{64\pi}\int\left\langle \theta(x)\theta(0)\right\rangle
_{\text{flat}}x^{4}d^{2}x\label{b1}%
\end{equation}
In an integrable model the flat space correlation functions in principle can
be constructed in terms of the form-factors of the operator $\theta$, which
are typically known exactly. However, even numerical calculation of the
three-point function (or multipoint ones for the next coefficients) in this
framework presents a separate problem.

Anyway, in the lack of any exact non-perturbative approach, a kind of
``experimental'' tool to measure the observables under consideration is very desirable.

\section{Schr\"odinger picture in sphere geometry}

In \cite{YuZ} the truncated conformal space (TCS) approach has been used to
evaluate numerically certain characteristics of 2D perturbed CFT's. In this
paper and a number of subsequent works \cite{TCS} it was demonstrated that TCS
is reasonably effective for many 2D models. In this section a way to apply
similar approach in the spherical geometry is discussed.

For our purpose it is convenient to map the projective coordinates $(z,\bar
z)$ on the sphere to the ``cylindric'' ones $(t,\sigma)$
\begin{equation}
z=\exp\xi=\exp(t+i\sigma)\label{tsigma}%
\end{equation}
where the sphere metric (\ref{sphere}) has the form
\begin{equation}
e^{\varphi(t,\sigma)}=\frac{R^{2}}{\cosh^{2}t}\label{phits}%
\end{equation}

Consider the space of states on a circle of fixed $\sigma$. We will be
interested in the ``time'' $t$ evolution of these states and, in the TCS
spirit, choose as the (time independent) basis the set of CFT states $\left|
a\right\rangle $ (here index $a$ runs over all the CFT states, primary and
descendent). For simplicity we suppose that the basis is orthonormal. Any
state (Schr\"odinger picture is implied where the states are time dependent)
\begin{equation}
\left|  \Psi(t)\right\rangle =\sum_{a}\Psi^{a}(t)\left|  a\right\rangle
\label{CFTbasis}%
\end{equation}
is described thus by the wave function $\Psi^{a}(t)$. Forgetting for the
moment the conformal anomaly (whose effect on the partition function is
summarized to the multiplier $(2R)^{c/3}$ in (\ref{ZR})) we consider the
following time dependent Hamiltonian
\begin{equation}
D(t)=D_{0}+g(t)V\label{Dt}%
\end{equation}
which generates translations in the $t$ direction. Here
\begin{equation}
D_{0}=L_{0}+\bar L_{0}\label{D0}%
\end{equation}
and $L_{0}$, $\bar L_{0}$ are the standard Virasoro generators acting on the
CFT states. The interaction part of the evolution is constructed as
\begin{equation}
V=\frac1{2\pi}\int_{0}^{2\pi}\phi(0,\sigma)d\sigma\label{V}%
\end{equation}
The time-dependent coupling constant $g(t)$ in (\ref{Dt}) reads in terms of
the ``effective coupling'' (\ref{h})
\begin{equation}
g(t)=-\frac h{(2\cosh t)^{2-2\Delta}}\label{gt}%
\end{equation}

Matrix elements $V\left|  a\right\rangle =\sum_{b}B_{a}^{b}\left|
b\right\rangle $ between CFT states are essentially the CFT structure
constants
\begin{equation}
B_{a}^{b}=\frac1{2\pi}\int_{0}^{2\pi}\left\langle f\right|  \phi
(0,\sigma)\left|  i\right\rangle d\sigma=C_{\phi\,a}^{b}\label{B}%
\end{equation}
They are constructed explicitly in solvable CFT models. The $t$-evolution of
the wave function is described by the time-dependent Schr\"odinger equation
\begin{equation}
-\frac d{dt}\left|  \Psi(t)\right\rangle =D(t)\left|  \Psi(t)\right\rangle
\label{Schr}%
\end{equation}
In the CFT basis we arrive at the infinite dimensional system of linear
differential equations
\begin{equation}
-\frac d{dt}\Psi^{a}(t)=\sum_{b}D_{b}^{a}(t)\Psi^{b}(t)\label{schr}%
\end{equation}

Different solutions to this system correspond to various operators placed at
the points $t=-\infty$ and $t=\infty$ (i.e., at the ``north'' and ``south''
poles of the sphere). In particular, the solution $\Psi_{\text{vac}}^{a}(t)$
determined by the initial condition
\begin{equation}
\Psi_{\text{vac}}^{a}(-\infty)=\delta_{I}^{a}\label{psivac}%
\end{equation}
where $a=I$ is the CFT state corresponding to the identity operator, describes
the state radiated by smooth (no field) north pole. The reduced partition
function (\ref{zh}) reads from this solution
\begin{equation}
z(h)=\lim_{t\rightarrow\infty}\Psi_{\text{vac}}^{I}(t)\label{zpsi}%
\end{equation}
It is easy to verify that the formal development (which implies that all
infinite sums over intermediate states are convergent) of the solution
$\Psi_{\text{vac}}$ in powers of $h$ results in the same perturbative series
(\ref{zn}), (\ref{zint}). From other components of the wave function one can
read-off the one-point functions at the south pole
\begin{equation}
(2R)^{2\Delta_{a}}\left\langle \Phi_{a}\right\rangle =\lim_{t\rightarrow
\infty}\frac{\Psi_{\text{vac}}^{a}(t)\exp(2\Delta_{a}t)}{\Psi_{\text{vac}}%
^{I}(t)}\label{Phia}%
\end{equation}
Here $\Phi_{a}$ is a scalar field corresponding to CFT state $\left|
a\right\rangle $ and $\Delta_{a}$ is its dimension. Of course, this limit is
in general divergent and requires standard UV renormalization, necessary to
define the perturbed fields (see e.g. \cite{LYcorr, desc}). This point will be
discussed in more details in \cite{one-point} while here we'll concentrate
only on the partition function, where the limit (\ref{zpsi}) is well defined.

The TCS idea is quite simple: truncate the infinite dimensional CFT space of
states up to certain maximal dimension and then treat the resulting
finite-dimensional problem (\ref{schr}) numerically. Previous TCS experience
shows that such procedure is often convergent (sometimes rather fast) with the
increase of the truncation dimension.

\section{Scaling Lee-Yang model}

Scaling Lee-Yang model (SLYM) \cite{SLYM} is often used as a testing tool for
various approximate approaches in 2D field theory. It is probably the simplest
(excluding, of course, the free field theories) example of perturbed CFT. The
model arises as a perturbation of the non-unitary CFT minimal model
$\mathcal{M}(2/5)$. This CFT is rational and contains only two primary fields,
the identity $I$ of dimension $0$ and the basic field $\phi=\Phi_{1,3}$ of
dimension $\Delta=-1/5$, the central charge being negative $c=-22/5$. The
basic operator product expansion reads
\begin{equation}
\phi(x)\phi(0)=(x\bar x)^{2/5}I+(x\bar x)^{1/5}C_{\phi\phi}^{\phi}%
\phi(0)+\ldots\label{ppp}%
\end{equation}
where the three-$\phi$ structure constant is purely imaginary $C_{\phi\phi
}^{\phi}=i\kappa$ and
\begin{equation}
\kappa=\left(  \frac{\sqrt{5}-1}2\right)  ^{1/2}\frac{\Gamma^{2}(1/5)}%
{5\Gamma(3/5)\Gamma(4/5)}=1.911312699\ldots\label{kappa}%
\end{equation}

The petrurbed model
\begin{equation}
\mathcal{A}_{\text{SLYM}}=\mathcal{A}_{\text{CFT}}+\frac{i\lambda}{2\pi}%
\int\phi(x)e^{\varphi(x)}d^{2}x\label{ASLYM}%
\end{equation}
is what is called the scaling Lee-Yang model\footnote{Notice that the coupling
$\lambda$ in (\ref{ASLYM}) is taken purely imaginary. This is nessessery to
make the perturbed theory real.}. In flat space-time ($\varphi\equiv0$) it is
integrable. Its particle content and factorized scattering theory are known
exactly \cite{CM}. The spectrum contains only one massive particle of mass $m$.

Integrability allows to obtain many exact results about SLYM. E.g., the
relation
\[
m=k\lambda^{5/12}
\]
between the mass $m$ and the coupling constant $\lambda$ is known
\cite{masscale}
\begin{equation}
k=\frac{4\sqrt{\pi}}{\Gamma(5/6)\Gamma(2/3)}\left[  \frac{\Gamma
(4/5)\Gamma(3/5)}{100\Gamma(1/5)\Gamma(2/5)}\right]  ^{5/24}%
=1.2288903248\ldots\label{mr}%
\end{equation}
The bulk vacuum energy is also found exactly \cite{TBA,DeVega}%

\begin{equation}
\mathcal{E}_{\text{vac}}=-\frac{m^{2}}{4\sqrt{3}}\label{ESLYM}%
\end{equation}
so that the dimensionless parameter $A$ in (\ref{Evac}) is
\begin{equation}
A_{\text{exact}}=\frac{k^{2}}{4\sqrt{3}}=0.2179745\ldots\label{A}%
\end{equation}

\section{TCS at low truncations}

In this section TCS for the Schr\"odinger equation (\ref{schr}) is applied to
the scaling Lee-Yang model. This model is especially convenient for TCS thanks
to the small number of primary fields and high degeneracy of modules at low
levels. This results in low-dimensional truncated spaces even for relatively
large truncation levels and makes the finite dimensional problem especially
simple for numerical treatment. E.g., truncation level 5 (which means the
maximum scale dimension in the truncated space $2\Delta_{\text{max}}=10$)
gives 17-dimensional space of states.

Up to level 5 the matrix elements $B_{i}^{f}$ can be found in \cite{YuZ}.
Truncated system (\ref{schr}) was solved numerically and the reduced partition
function $z(h)$ is evaluated as (\ref{zpsi}). In fig.1 the combination
\begin{equation}
\frac{Z_{\lambda}(R)}{Z_{0}}\exp(4\pi R^{2}\mathcal{E}_{\text{vac}}%
)=\Psi_{\text{vac}}^{I}(\infty)R^{-22/15}\exp(4\pi R^{2}\mathcal{E}%
_{\text{vac}})\label{Znum}%
\end{equation}
is plotted as a function of dimensionless variable $mR=(k/2)h^{5/12}$, where
$m$ is the mass of SLYM particle. Data for different truncation levels are
presented to illustrate the convergence.%

\begin{figure}
[tbh]
\begin{center}
\includegraphics[
height=3.7291in,
width=4.7893in
]%
{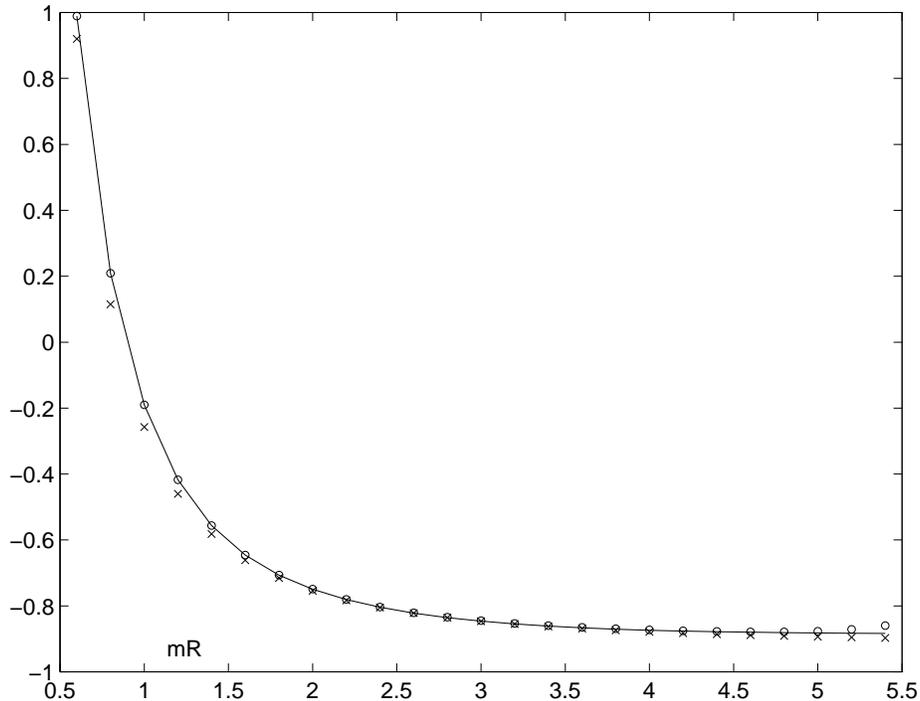}%
\caption{Combination (\ref{Znum}) for truncation levels 5 (full curve) and 4
(small circles). The IR fit with formula (\ref{z0TCS}) is plotted in crosses.}%
\label{fig1}%
\end{center}
\end{figure}

First interesting feature we observe is a zero of the partition function at
real positive $R=R_{0}$. Its position is estimated as
\begin{equation}
mR_{0}=0.88964\ldots\;\;\;\text{or \ \ \ }h_{0}=2.43083\ldots\label{h0}%
\end{equation}

At small $R$ the data are well fitted by the perturbative series (\ref{zn})
\begin{equation}
\frac{Z_{\lambda}(R)}{Z_{0}R^{-22/15}}=z(h)=1-\frac5{56}h^{2}-z_{3}%
h^{3}+\ldots\label{zhLY}%
\end{equation}
where $z_{3}$ can be computed numerically
\begin{align}
z_{3} &  =\frac{\kappa\Gamma(13/5)\Gamma(11/5)}{24\Gamma(24/5)}\int_{0}%
^{1}v^{1/5}dvF(-1/5,-1/5,1,v)F(12/5,13/5,24/5,1-v)\label{z3n}\\
&  =0.0229899\ldots\nonumber
\end{align}

In the relatively stable interval $2.0<mR<3.0$ the level 5 data were fitted,
according to (\ref{ZIR}), by the formula
\begin{equation}
\frac{Z_{\lambda}(R)\exp(4\pi R^{2}\mathcal{E}_{\text{vac}})}{Z_{0}}%
=Z_{\infty}\left(  1+\frac{b_{1}}{R^{2}}\right) \label{zinfb}%
\end{equation}
The best fit is achieved at
\begin{align}
Z_{\infty} &  =(-0.92\pm0.02)m^{-22/15}\label{z0TCS}\\
b_{1} &  =(-0.72\pm0.2)m^{-2}\nonumber
\end{align}
For dimensionless constants $z_{\infty}$ and $b_{1}$ in (\ref{zIR}) this
gives
\begin{align}
z_{\infty} &  =-1.25\pm0.03\label{Z0TCS}\\
a_{1} &  =-1.9\pm0.5\nonumber
\end{align}
These numbers enter the $h\rightarrow\infty$ asymptotic expansion
\begin{equation}
\log z(h)=\pi Ah^{6/5}+\log(2^{-22/15}z_{\infty})+\frac{11}{18}\log
h+a_{1}h^{-5/6}+a_{2}h^{-10/6}+\ldots\label{zLY}%
\end{equation}
where $A$ is given by (\ref{A}).

Analytic continuation for negative values of $h$ is obtained by
straightforward treatment of the truncated linear problem (\ref{schr}) with
negative $h$. The negative $h$ data show an oscillating pattern presented in
fig.2, where the values of $z(h)\exp\left(  \pi A\cos(\pi/6)(-h)^{5/6}\right)
(-h)^{-11/18}$ are plotted (see below for the asymptotic at $h\rightarrow
-\infty$) against $-h$.
\begin{figure}
[tbh]
\begin{center}
\includegraphics[
height=3.7291in,
width=4.7392in
]%
{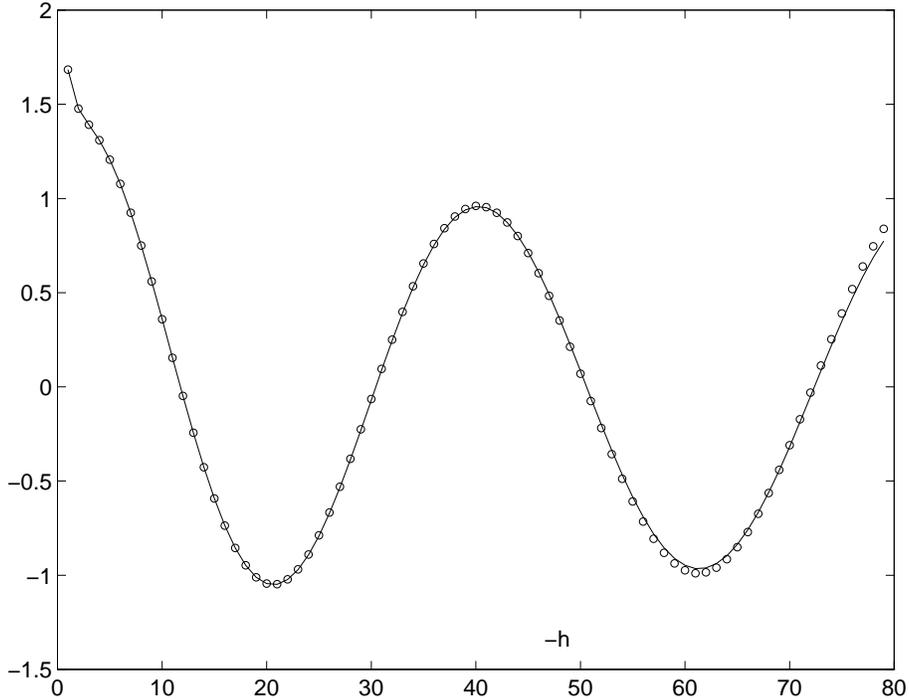}%
\caption{The negative $h$ data for $z(h)\exp\left(  \pi A\cos(\pi
/6)(-h)^{5/6}\right)  (-h)^{-11/18}$ at truncation levels 5 (continuous) and 4
(circles)}%
\label{fig3}%
\end{center}
\end{figure}

Numerical estimates for first few negative zeros $h_{n}$, $n=1,2,\ldots$ are
gathered in the first column of table \ref{table1}. It seems natural to expect
infinite number of zeros on the negative $h$ axis. Their positions are quite
important for as we'll argue below $z(h)$ is entire function of $h$ and
therefore essentially determined by the location of its zeros.

\begin{center}%
\begin{table}[tbp] \centering
\begin{tabular}
[c]{llll}%
$n$ & TCS & leading asymptotic & sum rules\\
$0$ & $2.43083$ &  & $2.43070$\\
$1$ & $-11.762$ & $-12.41$ & $-11.7731$\\
$2$ & $-30.439$ & $-30.66$ & $-30.2346$\\
$3$ & $-50.60$ & $-51.05$ & \\
$4$ & $-72.38$ & $-72.94$ & \\
$5$ &  & $-95.98$ & \\
$6$ &  & $-120.0$ & \\
$7$ &  & $-144.9$ &
\end{tabular}
\caption{Zeros of the partition function $z(h)$ estimated by TCS,
leading asymptotic (\ref{hn}) and sum rules(\ref{sumrules})\label{table1}}%
\end{table}
\end{center}

\section{Analytic considerations}

For any truncated finite dimensional linear problem (\ref{schr}),
(\ref{psivac}) and (\ref{zpsi}) the resulting reduced partition function
$z_{\text{trunc}}(h)$ is apparently an entire function of $h$. Let's suppose
that this property holds also for the exact function $z(h).$ We admit also
that all zeros of $z(h)$ are real, the first one $h_{0}$ being positive and
the rest $h_{n}$, $n=1,2,\ldots$ negative (and accumulating at $h=-\infty$
along the real axis). This implies in particular that the asymptotic
(\ref{zLY}) holds in the whole complex $h$-plane as $\left|  h\right|
\rightarrow\infty$, excluding the negative real axis $\arg h=\pm\pi$. This
analytic structure combined with the perturbative information collected in
(\ref{zh}) turns out to be quite restrictive.

Asymptotic (\ref{zLY}) allows to estimate the leading behavior of the
locations of zeros $h_{n}$ at $n\rightarrow\infty$%
\begin{equation}
-h_{n}=\left(  \frac2A\left(  n-\frac19\right)  +O(n^{-1})\right)
^{6/5}\label{hn}%
\end{equation}
In table \ref{table1} this asymptotic estimate is compared with the
approximate TCS data for first several zeros. The asymptotic does surprisingly
good even for the first negative zero $h_{1}$.

With (\ref{hn}) the series
\begin{equation}
\frac d{dh}\log z(h)=\sum_{n=0}^{\infty}\frac1{h-h_{n}}\label{ML}%
\end{equation}
is convergent and defines the logarithmic derivative through the positions of
zeros $h_{n}$. From (\ref{zhLY}) we have $\log z(h)=-\frac5{56}h^{2}%
-z_{3}h^{3}+\ldots$ and therefore $h_{n}$ satisfy the following sum rules
\begin{align}
\sum_{n=0}^{\infty}\frac1{h_{n}} &  =0\nonumber\\
\sum_{n=0}^{\infty}\frac1{h_{n}^{2}} &  =\frac5{28}\label{sumrules}\\
\sum_{n=0}^{\infty}\frac1{h_{n}^{3}} &  =3z_{3}=0.0689697\ldots\nonumber
\end{align}
A quick inspection of table \ref{table1} shows that the leading asymptotic
(\ref{hn}) agrees the TCS data with at least 1\% accuracy even for the forth
zero $h_{3}$. Let's take this asymptotic expression as the exact one and use
the sum rules to recalculate the first three zeros. This results in the
numbers presented in the forth column of table \ref{table1}. They are
impressively close to the ``experimental'' positions measured by TCS. This
numerical observation can be considered as a strong support of the suggested
analytic structure of $z(h)$.

Once all zeros are located with enough precision, the partition function can
be restored as the convergent Weierstrass product
\begin{equation}
z(h)=\prod_{n=0}^{\infty}\left(  1-\frac h{h_{n}}\right)  \exp\left(  \frac
h{h_{n}}\right) \label{W}%
\end{equation}
(due to (\ref{hn}) and (\ref{sumrules}) the exponential multiplier here is not
necessary and added to improve the convergence). The resulting function $z(h)$
based on the above approximation for the zeros $h_{n}$ is compared with level
5 TCS data for both positive and negative $h$ in figs. 3 and 4 respectively.%

\begin{figure}
[tbh]
\begin{center}
\includegraphics[
trim=0.000000in 0.000000in -0.003296in -0.001864in,
height=3.7291in,
width=4.7089in
]%
{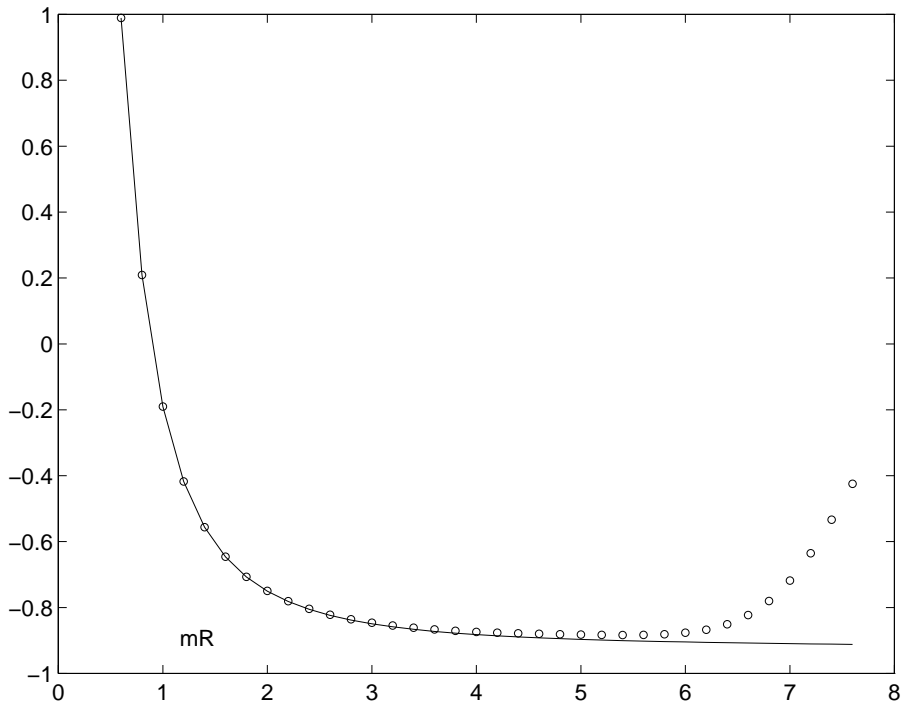}%
\caption{Weierstrass product (\ref{W}) for $Z_{\lambda}(R)\exp(4\pi
R^{2}\mathcal{E}_{\text{vac}})/Z_{0}$ based on the zeros in table \ref{table1}
(full curve) against the level 5 TCS data.}%
\label{fig4}%
\end{center}
\end{figure}

\begin{figure}
[tbh]
\begin{center}
\includegraphics[
trim=0.000000in 0.000000in 0.005246in -0.001864in,
height=3.7291in,
width=4.7694in
]%
{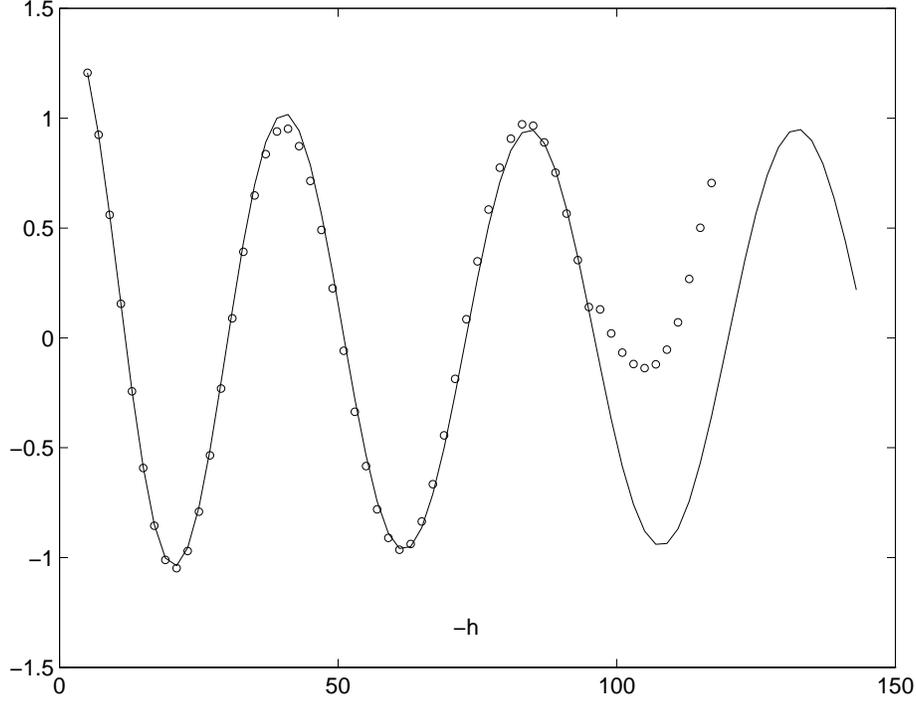}%
\caption{The same for negative values of the effective coupling constant $h$.
Values of $z(h)\exp\left(  \pi A\cos(\pi/6)(-h)^{5/6}\right)  (-h)^{-11/18}$
are plotted.}%
\label{fig5}%
\end{center}
\end{figure}

In principle one could improve the precision using next-to-leading correction
to the asymptotic (\ref{hn})
\begin{equation}
-h_{n}=\left(  \frac2A\left(  n-\frac19\right)  +\frac{a_{1}}{2\pi
(n-1/9)}+O(n^{-2})\right)  ^{6/5}\label{hn1}%
\end{equation}
with ``experimental'' value (\ref{Z0TCS}) for the coefficient $a_{1}$. Here I
prefer not to go along this line. One reason is the low precision in the
estimate (\ref{Z0TCS}). What is more important, the construction of $z(h)$ via
(\ref{W}) haven't used any numerical TCS data at all, the letter being only
the basis for certain hypotheses about the analytic structure. Moreover, the
only non-perturbative information (which in fact comes from the integrability
of SLYM in flat space-time) entering our above calculations is the exact
vacuum energy (\ref{A}). In the next section we'll see that even this income
can be omitted (for the price of some tolerable loss of precision) and $z(h)$
is restored rather accurately on the only ground of few first perturbative coefficients.

\section{Integrability off}

Let's artificially switch off the exact information (\ref{A}) and take this
number as an undeterminate parameter in the asymptotic expansion (\ref{zLY}),
together with $z_{\infty}$ and $a_{1,}a_{2},\ldots$. Of course, we continue to
keep in mind the analytic structure of $z(h)$ described above and denote the
zeros $h_{n}$, $n=0,1,2,\ldots$ as before. In particular the large $n$
asymptotic (\ref{hn}) for $h_{n}$ still holds.

To handle the zeros it is convenient to define (at $\operatorname*{Re}s>1$)
their zeta-function
\begin{equation}
\zeta_{\text{LY}}(s)=\frac{e^{5i\pi s/6}}{h_{0}^{5s/6}}+\sum_{n=1}^{\infty
}\frac1{(-h_{n})^{5s/6}}\label{zetaLY}%
\end{equation}
At $12/5>\operatorname*{Re}s>1$ it can be computed from the partition function
$z(h)$ as
\begin{equation}
\zeta_{\text{LY}}(s)=\frac{\sin(5\pi s/6)}\pi\int_{0}^{\infty}h^{-5s/6}%
\frac{d\log z(h-i0)}{dh}dh\label{zetaint}%
\end{equation}
where the contour of integration is shifted to agree the branch chosen in
(\ref{zetaLY}). Vise versa
\begin{equation}
\log z(h)=\int_{-i\infty}^{i\infty}\frac{\pi\zeta_{\text{LY}}(s)h^{5s/6}}%
{\sin(5\pi s/6)}\frac{ds}{2\pi is}\label{zinv}%
\end{equation}
where the integration contour goes to the left of the pole at $s=6/5$ (in fact
at $s=6/5$ there is no pole because $\zeta_{\text{LY}}(s)$ vanishes at this
point due to the sum rules, so that the first pole in the right half-plane
appears at $s=12/5$) and to the right of the pole at $s=1$. The following
properties of $\zeta_{\text{LY}}(s)$ are readily figured out:

\begin{enumerate}
\item $\zeta_{\text{LY}}(-6n/5)=0$ for $n=1,2,3,\ldots$. This is to avoid
wrong powers of $h$ in the asymptotic expansion (\ref{zLY}).

\item $\zeta_{\text{LY}}(s)$ has simple pole at $s=1$ with
$\operatorname*{res}_{s=1}\zeta_{\text{LY}}(s)=A/2$. This provides the correct
leading asymptotic in (\ref{zLY}).

\item $\zeta_{\text{LY}}(0)=11/18$ to ensure correct $\log h$ contribution in
the asymptotic, while the constant term requires

\item $\zeta_{\text{LY}}^{\prime}(0)=\dfrac56\log Z_{\infty}-\dfrac{11}9\log2$.

\item $\zeta_{\text{LY}}(s)$ has simple poles at all integer negative $s=-n $,
$n=1,2,\ldots$ and $\operatorname*{res}_{s=-n}\zeta_{\text{LY}}(s)=\pi
^{-1}n\sin(5\pi n/6)a_{n}$ where $a_{1},a_{2},\ldots$ are the subleading
coefficients in (\ref{zLY}).

\item  For all positive integer $k=1,2,\ldots$ we have by definition
$\zeta_{\text{LY}}(6k/5)=\sum_{n=0}^{\infty}(-h_{n})^{-k}$, i.e., the numbers
determined by the conformal perturbation theory. In particular (see
(\ref{sumrules}))
\begin{align}
\zeta_{\text{LY}}(6/5)=0\nonumber\\
\zeta_{\text{LY}}(12/5)=5/28\label{zetasum}\\
\zeta_{\text{LY}}(18/5)=-3z_{3}=-0.0689697\ldots\nonumber
\end{align}
\end{enumerate}

The simplest approximation here is to replace all $h_{n}$ for $n=2,3,\ldots$
by the leading asymptotic (\ref{hn}) while leaving $A$, $h_{0}$ and $h_{1}$ as
unknowns. In other words we approximate $\zeta_{\text{LY}}(s)$ as
\begin{equation}
\zeta_{\text{LY}}^{\text{(0)}}(s)=\frac{e^{5i\pi s/6}}{h_{0}^{5s/6}}%
+\frac1{(-h_{1})^{5s/6}}+\left(  \frac A2\right)  ^{s}\zeta
(s,17/9)\label{zetaLY0}%
\end{equation}
where $\zeta(s,a)=\sum_{n=0}^{\infty}(n+a)^{-s}$ is the usual Riemann
zeta-function. This expression identically satisfies property 3 while the sum
rules (\ref{sumrules}) can be considered as a system of equations for
undeterminate parameters $A$, $h_{0}$ and $h_{1}$. There are two real
solutions. The first one
\begin{align}
A &  =0.218156\ldots\nonumber\\
h_{0} &  =2.43068\ldots\label{first}\\
h_{1} &  =-11.7540\ldots\nonumber
\end{align}
is really impressive when compared with the exact value (\ref{A}) for $A$ and
the ``experimental'' results for $h_{n}$ from table \ref{table1}. It's not
clear if this remarkable precision is accidental. I'd like to remind that it's
not a first instance of miraculous numerical success while applying different
approximations to this particular perturbed CFT \cite{YuZ, LYcorr}.

With these data in hand we can estimate also the constant $z_{\infty}$
evaluating $\zeta_{\text{LY}}^{\text{(0)}\prime}(0)$%
\begin{equation}
z_{\infty}=-\frac{2^{22/15}\Gamma(17/9)^{6/5}}{h_{0}h_{1}(A/2)^{5/3}%
(2\pi)^{3/5}}=-1.22491\ldots\label{Zinf0}%
\end{equation}
again in reasonable agreement with the TCS analysis.

Approximation (\ref{zetaLY0}) seems to be quite accurate in the right
half-plane of $s$. However it has wrong analytic properties in the left
half-plane. For example it has no pole at $s=-1$ (as well as at all $s$
negative integer) and doesn't vanish at $s=-6n/5$, $n=1,2,\ldots$. E.g.,
\begin{equation}
\zeta_{\text{LY}}^{\text{(0)}}(-6/5)=-h_{0}-h_{1}+\left(  \frac2A\right)
^{6/5}\zeta(-6/5,17/9)=-3.28339\ldots\label{defect}%
\end{equation}
One can try to cure the situation near $s=-1$ introducing a pole at $s=-1$ by
hand, i.e., taking say (this ansatz is obviously inspired by eq.(\ref{hn1}))
\begin{equation}
\zeta_{\text{LY}}^{\text{(1)}}(s)=\frac{e^{5i\pi s/6}}{h_{0}^{5s/6}}%
+\frac1{(-h_{1})^{5s/6}}+\left(  \frac A2\right)  ^{s}\zeta(s,17/9)-\frac
{a_{1}}{2\pi}s\left(  \frac A2\right)  ^{s+1}\zeta(s+2,17/9)\label{zetaLY1}%
\end{equation}
with a new undeterminate parameter $a_{1}$. Then, with the previous values of
$A$, $h_{0}$ and $h_{1}$ we find that $\zeta_{\text{LY}}^{\text{(1)}}(-6/5)=0$
if
\begin{equation}
a_{1}\approx-2.060\label{a1}%
\end{equation}
in tolerable agreement with (\ref{Z0TCS}). It can be verified that this
correction doesn't violate essentially the sum rules (\ref{zetasum}) while
$z_{\infty}$ is slightly changed $z_{\infty}=-1.26190\ldots$.

In principle one can continue the procedure and cure the value of
$\zeta_{\text{LY}}(-12/5)$ by adding more terms and trying to correct the
analytic structure near $s=-2$, $s=-3$ etc. Due to the asymptotic character of
the expansion (\ref{zLY}), such iterative procedure is hardly convergent.
Nevertheless the estimates of the unknown parameters $A$, $h_{0}$, $h_{1}$ and
$a_{1}$ looks quite reasonable. I don't know yet any explanation for such
success. Probably it's accidental.

The second real solution to (\ref{zetasum}) is numerically quite close to the
first one
\begin{align}
A &  =0.305435\ldots\nonumber\\
\;h_{0} &  =2.44244\ldots\;\;\;\label{second}\\
h_{1} &  =-12.5161\ldots\nonumber
\end{align}
The estimate (\ref{Zinf0}) gives now $z_{\infty}=-0.64821\ldots$. It remains a
question whether this solution is an artefact of the approximate approach or
we must take it seriously and look for a proper interpretation. Let me note
that with the second value of $A$ (\ref{second}) the ``jump'' between the
asymptotic estimate (\ref{hn}) of $h_{n}$ and $h_{1}$ from (\ref{second}) is
rather big. This makes the whole reasoning above very questionable.

\section{Remarks}

\begin{itemize}
\item  The most interesting conclusion we're led by both TCS experiment and
analytic considerations (in the framework of SLYM) is that the spherical
partition function of a perturbed CFT may have quite specific analytic
properties in the coupling constant (\ref{h}). Are these properties are
typical for any perturbed CFT on a sphere? Exactly solvable example of free
massive fermion (see Appendix A.1, where the partition functions of free
massive fields are presented explicitly) shows very similar analytic
structure. Combining these two examples one might conclude that the answer is
probably yes. However, the free massive boson (Appendix A.2) gives an
immediate counterexample. Analytic structure is more complicated, the
partition function developing an infinite sequence of brunch points etc.
Probably the simple analytic picture is characteristic for rational perturbed
CFT's. It remains to be answered if integrability of the corresponding model
in flat plays any role in this analytic structure.

\item  Consider the derivative $z^{\prime}(h)=dz(h)/dh$, which is again
supposed to be an entire function of $h$ with the asymptotic at $\left|
z\right|  \rightarrow\infty$, $-\pi<\arg z<\pi$
\begin{equation}
z^{\prime}(h)\sim\exp\left(  \pi Ah^{5/6}+\frac49\log h+\text{const}%
+...\right) \label{zprim}%
\end{equation}
Denote its zeros $h_{n}^{\prime}$. Perturbative expansion (\ref{zhLY}) gives
exactly the first zero $h_{0}^{\prime}=0$. From (\ref{zprim}) the large $n$
asymptotic of $h_{n}^{\prime}$ is read off
\begin{equation}
-h_{n}^{\prime}=\left(  \frac2A\left(  n+\frac1{18}\right)  +O(n^{-1})\right)
^{6/5}\label{hnprim}%
\end{equation}
In particular, the convergent product
\begin{equation}
z^{\prime}(h)=-\frac5{28}h\prod_{n=1}^{\infty}\left(  1-\frac h{h_{n}^{\prime
}}\right) \label{Wprim}%
\end{equation}
relates $z^{\prime}(h)$ to the positions of its zeros. The last known term in
the series (\ref{zhLY}) results in the sum rule for $h_{n}^{\prime}$
\begin{equation}
\sum_{n=1}^{\infty}\frac1{h_{n}^{\prime}}=-\frac{84}5z_{3}\label{sumprim}%
\end{equation}
Take the leading asymptotic (\ref{hnprim}) as exact for $n\geq1$. The sum rule
becomes
\begin{equation}
\ \ \left(  \frac A2\right)  ^{6/5}\sum_{n=1}^{\infty}\frac1{(n+1/18)^{6/5}%
}=\frac{84}5z_{3}=0.3862303\label{Aeq}%
\end{equation}
and gives for $A$
\begin{equation}
A=0.218767...\label{Aprim}%
\end{equation}
This accurate result is again remarkable, especially in view of the amazingly
simple way it is obtained. Also, unlike more complicated system in sect.9,
this way gives a unique solution.

\item  Note, that the derivative $z^{\prime}(h)$ is related to the
(unnormalized) one-point function of the perturbing operator $\phi$%
\begin{equation}
z(h)(2R)^{-2/5}\left\langle \phi\right\rangle =-2z^{\prime}(h)\label{phivev}%
\end{equation}
In the TCS approach
\begin{equation}
z(h)(2R)^{-2/5}\left\langle \phi\right\rangle =\lim_{t\rightarrow\infty}%
\Psi_{\text{vac}}^{\phi}(t)\exp(-2t/5)\label{Psiphi}%
\end{equation}
(no subrtactions are required for this one-point function). Apparently, for
any finite-dimensional TCS problem this is again an entire function of $h.$ It
seems natural to generalize the observed analytic structure for arbitrary
unnormalized one-point function
\begin{equation}
z(h)(2R)^{2\Delta_{a}}\left\langle \Phi^{a}\right\rangle =\lim_{t\rightarrow
\infty}\left(  \Psi_{\text{vac}}^{a}(t)\exp(2\Delta_{a}t)-\text{subtractions}%
\right) \label{vPhia}%
\end{equation}
and then use it to get more information on the grounds of perturbation theory
and analytic properties. These attempts will be reported in a separate
publication \cite{one-point}.

\item  It would be interesting to compare both the experimental (\ref{z0TCS})
and ``theoretical'' (\ref{a1}) estimates of the first supleading IR correction
to the partition function with the prediction (\ref{b1}) evaluated with the
use of exact form-factors of $\theta$ \cite{LYcorr}.

\item  TCS approach as described above proves to provide reasonable
experimental data for the perturbed CFT on the sphere. It is well known
however, that this method works well mainly for perturbed rational CFT's, and
moreover, when the conformal perturbative integrals are convergent. In
particular, many interesting models where the dimension of perturbing operator
is close to $1$, or with asymptotically free marginal perturbation are
completely unaccessible by this numerical scheme. A suitable approach which
would allow to overcome this restriction still remains to be developed.

\item  Nevertheless, there are still many perturbed CFT's (like the sin-Gordon
model with sufficiently small $\beta$) where TCS must perform reasonably and
the hypotheses proposed in this article can be checked against the
experimental data. Work in this direction is in progress.

\end{itemize}
\vspace{8cm}
\appendix{\Large \textbf{Appendix}}

\section{Free fields on a sphere}

\subsection{Majorana fermion}

Free massive Majorana fermion in curved geometry with conformally flat metric
(\ref{isothermal}) is defined by the action
\begin{align}
\mathcal{A}_{\text{ferm}} &  =\frac1\pi\int d^{2}x\left[  \psi\bar\partial
\psi+\bar\psi\partial\bar\psi+ime^{\varphi/2}\bar\psi\psi\right]
\label{Aferm}\\
\  &  =\frac1\pi\int d^{2}x\left[  \psi\bar\partial\psi+\bar\psi\partial
\bar\psi\right]  +\frac m{2\pi}\int\varepsilon(x)e^{\varphi(x)}d^{2}x\nonumber
\end{align}
where $m$ is the mass of the fermion, which is just the coupling constant in
the framework of perturbed CFT, while the energy density
\begin{equation}
\varepsilon=2ie^{-\varphi/2}\bar\psi\psi\label{eps}%
\end{equation}
plays the role of perturbing operator. This action leads to the Dirac
equations of motion
\begin{align}
2\bar\partial\psi &  =ime^{\varphi/2}\bar\psi\label{Dirac}\\
2\partial\bar\psi &  =-ime^{\varphi/2}\psi\nonumber
\end{align}

We're interested in the partition function $Z_{m}^{\text{(f)}}(R)$ of this
field theory on the sphere of radius $R.$ Obviously its derivative in $m$ is
determined by the expectation value of $\varepsilon$. By symmetry
$\left\langle \varepsilon(x)\right\rangle $ is independent on $x$ and
\begin{equation}
\frac{d\log Z_{m}^{\text{(f)}}(R)}{dm}=-2R^{2}\left\langle \varepsilon
\right\rangle \label{Zeps}%
\end{equation}
To evaluate $\left\langle \varepsilon\right\rangle $ consider the two-point
functions of the fermionic fields. For the symmetry reasons
\begin{align}
\left\langle \psi(z,\bar z)\psi(0)\right\rangle  &  =\frac1zf(z\bar
z)\label{psipsi}\\
\left\langle \bar\psi(z,\bar z)\psi(0)\right\rangle  &  =-ig(z\bar z)\nonumber
\end{align}
Equations of motion (\ref{Dirac}) result in the following simple system
\begin{align}
(1-t)f^{\prime}(t) &  =rg(t)\label{fdiff}\\
tg^{\prime}(t) &  =rf(t)\nonumber
\end{align}
where $r=mR$ and
\begin{equation}
t=\frac{z\bar z}{(1+z\bar z)}=\sin^{2}\left(  \frac s{2R}\right)
\label{geodesic}%
\end{equation}
is related to the geodesic distance $s$ between $0$ and $z.$ Suitable
solution, which is regular at $t=1$ (i.e., at the south pole of the sphere
$z=\infty$) and properly normalized at $t=0$ reads
\begin{align}
f(z\bar z) &  =\frac{\Gamma(1+ir)\Gamma(1-ir)}2F\left(  ir,-ir,1,(1+z\bar
z)^{-1}\right) \label{fg}\\
g(z\bar z) &  =-\frac{r\Gamma(1+ir)\Gamma(1-ir)}{2(1+z\bar z)}F\left(
1+ir,1-ir,2(1+z\bar z)^{-1}\right) \nonumber
\end{align}
Taking the limit $z\bar z\rightarrow0$ one can read-off the expectation value
$\left\langle \varepsilon(0)\right\rangle $ at the north pole $z=0$%
\begin{align}
\left\langle \varepsilon\right\rangle  &  =2g\left(  \epsilon^{2}%
e^{-\varphi(0)}\right)  e^{-\varphi(0)/2}\label{epsv}\\
\  &  =\frac m2\left(  \log\frac{\epsilon^{2}}{4R^{2}}+\psi(1+ir)+\psi
(1-ir)-2\psi(1)\right) \nonumber
\end{align}
where $\psi(x)=\Gamma^{\prime}(x)/\Gamma(x)$ and $\epsilon$ is the invariant
UV cutoff. Integrating now (\ref{Zeps}) and taking into account the conformal
anomaly (\ref{ZR}) one finds
\begin{equation}
\frac{Z_{m}^{\text{(f)}}(R)}{Z_{0}^{\text{(f)}}R^{1/6}}=\exp\left(
-r^{2}\left(  \log\dfrac\epsilon{2R}+C+1\right)  \right)  \frac{\exp
(2\zeta^{\prime}(-1))}{\Gamma_{2}(1+ir|1,1)\Gamma_{2}(1-ir|1,1)}\label{Zferm}%
\end{equation}
where $C$ is the Euler's constant and $\Gamma_{2}(x|1,1)$ is a particular case
of the Barnes double gamma function \cite{Barnes}
\begin{equation}
\log\Gamma_{2}(x|\omega_{1},\omega_{2})=\frac d{ds}\left.  \sum_{m,n=0}%
^{\infty}(x+m\omega_{1}+n\omega_{2})^{-s}\right| \label{dgamma}%
\end{equation}

Ferimionic partition function $Z_{m}^{\text{(f)}}(R)$ is an entire function of
$r$. It has zeros at $r=\pm i$, $\pm2i$, $\pm3i\ldots$ of multiplicities $1$,
$2$, $3\ldots$ respectively, which are of course related to the zero modes of
the spherical Dirac operator at these values of $mR$.

At small $R$ this partition function develops as follows
\begin{equation}
\log\frac{Z_{m}^{\text{(f)}}(R)}{Z_{0}^{\text{(f)}}}=\frac16\log R-r^{2}%
\log\frac\epsilon{2R}+\sum_{k=1}^{\infty}\frac{(-)^{k}\zeta(2k+1)}%
{k+1}r^{2k+2}\label{ZfermUV}%
\end{equation}
The large $R$ behavior is determined by the asymptotic expansion
\begin{equation}
\frac{Z_{m}^{\text{(f)}}(R)}{Z_{0}^{\text{(f)}}}\sim m^{-1/6}z_{\infty
}^{\text{(f)}}\exp\left(  -4\pi R^{2}\mathcal{E}_{\text{vac}}^{\text{(f)}%
}+\frac1{120m^{2}R^{2}}+\frac1{504m^{4}R^{4}}\ldots\right) \label{ZfermIR}%
\end{equation}
where
\begin{equation}
\mathcal{E}_{\text{vac}}^{\text{(f)}}=\frac{m^{2}}{4\pi}\left(  \log
\frac{m\epsilon}2+C-\frac12\right) \label{Eferm}%
\end{equation}
is the bulk vacuum energy of the free fermion in flat space-time and
\begin{equation}
z_{\infty}^{\text{(f)}}=\exp(2\zeta^{\prime}(-1))=0.718318\ldots\label{Zinff}%
\end{equation}

\subsection{Massive boson}

The action of free massive boson $\phi$ in curved geometry reads
\begin{equation}
\mathcal{A}_{\text{boson}}=\frac1{4\pi}\int d^{2}x\left[  (\partial_{a}%
\phi)^{2}+m^{2}e^{\varphi}\phi^{2}\right] \label{Aboson}%
\end{equation}
and lead to the free field equation of motion
\begin{equation}
4\partial\bar\partial\phi=m^{2}e^{\varphi}\phi\label{KG}%
\end{equation}

Again, the derivative in $m^{2}$ of the bosonic partition function
$Z_{m}^{\text{(b)}}(R)$ is expressed in terms of $\left\langle \phi
^{2}\right\rangle $ on the sphere
\begin{equation}
\frac{d\log Z_{m}^{\text{(b)}}(R)}{dm^{2}}=-\frac1{4\pi}\int\left\langle
\phi^{2}\right\rangle e^{\varphi}d^{2}x=-R^{2}\left\langle \phi^{2}%
\right\rangle \label{dZbdm2}%
\end{equation}
From (\ref{KG}) it follows that the two-point function $G(z\bar
z)=\left\langle \phi(z,\bar z)\phi(0)\right\rangle $ satisfies hypergeometric
differential equation
\begin{equation}
t(1-t)G_{tt}+(1-2t)G_{t}-m^{2}R^{2}G=0\label{bdiff}%
\end{equation}
where $t$ is the same as in (\ref{geodesic}). Relevant solution, regular at
the ``south pole'' $z=\infty$ and properly normalized at $z\rightarrow0$, has
the form
\begin{equation}
G(z\bar z)=\frac{\Gamma(1/2+is)\Gamma(1/2-is)}2F\left(
1/2+is,1/2-is,1,(1+z\bar z)^{-1}\right) \label{bhyper}%
\end{equation}
where
\begin{equation}
s=\left(  m^{2}R^{2}-1/4\right)  ^{1/2}\label{s}%
\end{equation}
Considering the subleading term in $z\rightarrow0$ asymptotic of this function
we find
\begin{align}
\left\langle \phi^{2}\right\rangle  & =G\left(  \epsilon^{2}e^{-\varphi
(0)}\right) \label{vphi2}\\
\  & =-\frac12\left(  \log\frac{\epsilon^{2}}{4R^{2}}+\psi(1/2+is)+\psi
(1/2-is)-2\psi(1)\right) \nonumber
\end{align}
where again the UV cutoff $\epsilon$ appears explicitly.

Integration of (\ref{dZbdm2}) is not that straightforward as in the fermionic
case. Expectation value (\ref{vphi2}) is singular at $s=\pm i/2$, i.e, at
$m=0$. This is of course a manifestation of the zero mode of (\ref{KG}) at
$m=0$. It leads to an extra multiplier $(mR)^{-1}$ (in addition to the
conformal anomaly (\ref{ZR})) in the small $R$ behavior of the partition
function. Normalization of $Z_{m}^{\text{(b)}}(R)$ becomes ambiguous even if
$Z_{0}^{\text{(b)}}$ is fixed in some way. Below we adopt the normalization
where $Z_{m}^{\text{(b)}}(R)/Z_{0}^{\text{(b)}}\sim R^{1/3}/(mR)$ at
$R\rightarrow0$. With this convention
\begin{align}
& \frac{Z_{m}^{\text{(b)}}(R)}{Z_{0}^{\text{(b)}}R^{1/3}}=\label{Zbdgamma}\\
& \exp\left(  m^{2}R^{2}\left(  \log\frac\epsilon{2R}+C+1\right)  \right)
\frac{\exp(-2\zeta^{\prime}(-1))\Gamma_{2}(3/2+is|1,1)\Gamma_{2}%
(3/2-is|1,1)}{(2\cosh(\pi s))^{1/2}}\nonumber
\end{align}
Bosonic partition function has brunch points of order $-1/2$ at $s=\pm i/2$,
of order $-3/2$ at $s=$ $\pm3i/2$, of order $-5/2$ at $s=\pm5i/2$ etc. These
singularities are caused by the zero modes of (\ref{KG}). In terms of the
variable $m^{2}R^{2}$ they turn to similar brunch points at $m^{2}%
R^{2}=-n(n+1)$, $n=0,1,2,\ldots$

Small $R$ behavior reads
\begin{equation}
\frac{Z_{m}^{\text{(b)}}(R)}{Z_{0}^{\text{(b)}}}\sim\frac{R^{1/3}}{mR}%
\exp\left(  \frac{m^{2}R^{2}}2\left(  \log\frac{\epsilon^{2}}{4R^{2}%
}+1\right)  +\frac14m^{4}R^{4}-\frac{\zeta(3)-1}3m^{6}R^{6}+\ldots\right)
\label{ZbUV}%
\end{equation}
while at $R\rightarrow\infty$%
\begin{equation}
\frac{Z_{m}^{\text{(b)}}(R)}{Z_{0}^{\text{(b)}}}=m^{-1/3}z_{\infty
}^{\text{(b)}}\exp\left(  -4\pi R^{2}\mathcal{E}_{\text{vac}}^{\text{(b)}%
}+\frac1{30m^{2}R^{2}}+\frac2{315m^{4}R^{4}}\ldots\right) \label{ZbIR}%
\end{equation}
where
\begin{equation}
\mathcal{E}_{\text{vac}}^{\text{(b)}}=-\frac{m^{2}}{4\pi}\left(  \log
\frac{m\epsilon}2+C-\frac12\right) \label{Evboson}%
\end{equation}
is the bulk vacuum energy of the massive free boson in flat, and
\begin{equation}
z_{\infty}^{\text{(b)}}=\exp\left(  1/4-2\zeta^{\prime}(-1)\right)
=1.78754\ldots\label{Zinfb}%
\end{equation}
\textbf{Acknowledgments.}

Al.Z thanks A.Zamolodchikov for important comments. Discussions with V.Fateev
and G.Mussardo were also of much help. The work was supported by EU under the
contract ERBFMRX CT 960012.

\end{document}